# Spin photovoltaic cell

Federico Bottegoni,[1]* Michele Celebrano,[1] Monica Bollani,[2] Paolo Biagioni,[1] Giovanni Isella,[1] Franco Ciccacci,[1] and Marco Finazzi[1]

[1] *LNESS-Dipartimento di Fisica, Politecnico di Milano, Piazza Leonardo da Vinci 32, 20133 Milano, Italy.*
[2] *CNR-IFN and LNESS, via Anzani 42, 22100 Como, Italy.*

* *e-mail* federico.bottegoni@polimi.it

*Spintronics, which aims at exploiting the spin degree of freedom of carriers inside electronic devices, has a huge potential for quantum computation and dissipationless interconnects. Ideally, spin currents in spintronic devices should be powered by a "spin voltage generator" able to drive spins out of equilibrium and produce two spatially well-separated populations with opposite spin orientation. Such a generator should work at room temperature, be highly integrable with existing semiconductor technology, and work with neither ferromagnetic materials nor externally applied magnetic fields. We have matched these requirements by realizing the spintronic equivalent of a photovoltaic cell. While the latter spatially separates photoexcited electron and hole charges, our device exploits circularly polarized light to produce two spatially well-defined electron populations with opposite spin. This is achieved by modulating the phase and amplitude of the light wavefronts entering a semiconductor (germanium) with a patterned metal overlayer (platinum). This allows creating a light diffraction pattern with spatially-modulated chirality inside the semiconductor, which locally excites spin-polarized electrons thanks to electric dipole selection rules.*

The integration of spintronics and photonics represents one of the most challenging tasks in modern optoelectronics[1],[2],[3],[4]. In this context germanium plays a key role, due to the large spin orbit coupling and electron spin lifetime [5],[6], simple integrability with Si-based electronic platforms, and optical properties matching the conventional telecommunication window [7],[8].In the process known as *optical orientation*[2], dipole selection rules, for direct optical transitions with circularly polarized light, allow the excitation of an electron population in the Ge conduction band with a spin polarization $P = (n\uparrow - n\downarrow)/(n\uparrow + n\downarrow)$ up to 50% [9],[10], where $n\uparrow_{(\downarrow)}$ are the up (down) spin densities (as referred to a proper quantization axis). An advantage of optical orientation is that neither ferromagnetic materials nor externally applied magnetic fields are needed to obtain spin-polarized electrons. However, the exploitation of optical orientation to excite spatially separated carriers with opposite spin and create spin accumulation in spintronic devices would require the spatial tuning of the light circular polarization over the device, a contrivance that, up to now, has not been deemed viable. Moreover, most spintronic devices require a carrier spin polarization in the plane of the device [11], which can be achieved by optical orientation only with grazing incidence illumination.

We solve these problems by spatially modulating the phase and amplitude of the electromagnetic wave-fronts impinging at normal incidence on the semiconductor sample by patterning a thin overlayer deposited onto the semiconductor surface. This method allows creating a field distribution inside the semiconductor with spatially-engineered chirality with respect to an *in-plane* quantization axis *y*, i.e. in the direction *perpendicular* to the wave vector of the illuminating light. This is associated to the Stokes parameter $c_y = 2\text{Im}(E_x E_z^*)$ [12], being $E_x$ and $E_z$ the (complex) components of the electric field along the *x* and *z* axes forming with *y* a Cartesian reference system where *z* is perpendicular to the semiconductor surface. The simplest configuration consists of a nonmagnetic metal stripe deposited on the semiconductor surface. Figure 1a presents a pictorial scheme of the structure and the simulated $c_y$ parameter when the stripe is illuminated at normal incidence with a circularly-polarized plane wave and a photon energy at the semiconductor direct band gap ($E_d = 0.8$ eV for Ge at room temperature). The scattering by the stripe creates a local field with a $c_y$ distribution (see Fig. 1a) that simultaneously excites two spatially separated populations of spin-oriented electrons with opposite spin projections *in the sample plane*; a similar spatial modulation can be retrieved considering the $c_x = 2\text{Im}(E_y E_z^*)$ parameter. This effect does not depend on the particular choice of the semiconductor/patterning material combination, but rather on the geometry of the patterning and on the dipole selection rules for the semiconductor. Notably, this concept can be confined to sub-micrometer scales (see below) and extended to a wide range of architectures and different semiconductor/patterning material combinations. Moreover, the generation of the two opposite spin densities is not limited by the spin diffusion length of the electrons along the chosen quantization axis, as it happens for electrically-induced edge spin accumulation [13]: in the proposed device, the spatial scale of the two generation profiles in the semiconductor is only driven by the geometry of the patterning.

The device proposed in Fig. 1a has been realized by uniformly evaporating a 4 nm-thick Pt layer on top of an As-doped Ge substrate (5×5 mm$^2$, $N_d = 2\times10^{16}$ cm$^{-3}$). A set of Pt stripes (length 20 μm, height 20 nm, width 2 μm, separated by 3.6 μm) has been obtained by electron beam lithography followed by a second evaporation step. Circularly polarized light ($hv = 0.8$ eV, resonant with the direct Ge bandgap) excites spin-polarized electron-hole pairs at the Γ point of the Ge Brillouin zone with a generation rate that is proportional to $c_y$. The spin-oriented photo-excited electrons lie at the bottom of the conduction band at Γ. Within ~300 fs they are scattered to the L minima, partially maintaining the initial spin orientation [6]. Since the electron spin lifetime in bulk Ge is in the nanoseconds range [5], whereas the hole spin lifetime is much shorter (few hundreds of femtoseconds [14]), the spatially modulated spin density can be assumed to be mainly due to electrons. We find that, with our geometry, the spin-polarization of the photo-excited electrons is projected into the sample plane with high efficiency compared with non-patterned devices exploiting optical spin injection at grazing incidence [15]. Indeed, we have calculated that the maximum $c_y$ value that can be obtained in a Ge sample covered with a uniform 4 nm-thick Pt layer is about one half of the maximum value obtained with the present setup.

Pt has been chosen as patterning material because, thanks to its high atomic number, it effectively functions as a spin detector. Spin accumulation can be revealed since it induces a pure spin current diffusing in Ge along the L states towards the Pt/Ge interface [15]. Injection of a spin current density **J**$_s$ into the Pt layer gives origin to the inverse spin-Hall effect (ISHE) via spin-dependent scattering with Pt nuclei (see Fig. 1c), resulting in the generation of an electromotive field **E**$_{ISHE}$ [16]

$$\mathbf{E}_{ISHE} = D_{ISHE}\, \mathbf{J}_s \times \mathbf{u_P}, \qquad (1)$$

where $D_{ISHE}$ is a coefficient that represents the ISHE efficiency and **u**$_P$ is the unit vector corresponding to the direction of the spin polarization **P**. We can then measure the electromotive force $\Delta V_{ISHE} \approx E_{ISHE} \cdot d$ ($d$ being the lateral size of the region from which electrons diffuse into the Pt overlayer), generated between two Au contacts evaporated at the edges of the Pt layer (see Fig. 1c and 1d). Due to the position of the contacts (see Fig. 1c), $\Delta V_{ISHE}$ is determined by the **E**$_{ISHE}$ component pointing along $x$. Since **J**$_s$ is parallel to the $z$ direction, our setup is thus sensitive to the $y$-component of **P**.

By using uniform illumination, as shown in Fig. 1a, the opposite spin-polarized electron populations generated at the two edges of the stripe provide spin-current densities that counterbalance each other, giving rise to a null $\Delta V_{ISHE}$. Therefore, to separately visualize each spin-polarized population, one needs to break the illumination symmetry. A diffraction-limited confocal setup (numerical aperture 0.7, giving a full-width-half-maximum beam size $w \approx 1.5$ µm) has been exploited to selectively generate the desired spin density at each edge of the Pt stripe (see Fig. 1d). At the same time, the optical reflectivity signal of the set of Pt stripes has been monitored by an InGaAs detector (see Fig. 2a-b). All measurements have been performed at room temperature.

The 2D scan of Fig. 2c clearly shows a strong $\Delta V_{ISHE}$ signal changing sign across the stripes. When the Gaussian beam illuminates one of the ridge edges along the $y$-axis, the spin polarization due to the photo-generated electrons in the Ge wafer dramatically increases. A reversed effect is observed on the opposite stripe edge. Asymmetries in the absolute value of the profile has to be ascribed to photoelectric effects associated with a residual light intensity modulation. Such effects induce a variation of the instrumental background at correspondence with the stripes, which have a different reflectivity than the surrounding film. Slight aberrations of the light beam might also contribute to these asymmetries.

By applying a one-dimensional spin drift-diffusion model, and reminding that, at $hv = E_d$, the initial electron spin-polarization along the $y$ axis is at most 50%, we estimate that the majority of the photoexcited electrons enter the Pt overlayer before depolarizing. This result stems from the light absorption length in Ge (which is about 2.2 μm at $hv = 0.8$ eV) being comparable with the spin diffusion length $\Lambda_{Ge} \approx 2.5$ μm of the conduction band electrons along the L states[5],[15]. Notably, the maximum absolute value of the differential voltage signal $\Delta V_{ISHE}$ ($\Delta V_{ISHE}$ = 2.4 mV at $W$ = 150 mW), when properly normalized to account for the different experimental set-ups, is similar to the value measured on a device exploiting optical orientation at grazing incidence [15], indicating that the spin polarization densities obtained in the two cases are comparable. This gives an experimental confirmation of the efficiency of the present configuration, as also obtained in the simulations discussed above yielding large $c_y$ values.

Numerical simulations allows us to correlate the experimental $\Delta V_{ISHE}$ map with the spin density in the $y$-direction (parallel to the stripe short axis) of photogenerated electrons in the Ge wafer, which is proportional to the Stokes parameter $c_y$. Figure 3a shows a map of $c_y$, calculated when the light beam is centered on one of the edges of the Pt stripe. The resulting interference pattern indicates a net positive $c_y$ value, which corresponds to the detection of an ISHE signal in the experimental geometry of Fig. 1d. Figure 3b shows the comparison between the $\Delta V_{ISHE}$ signal and the calculated $c_y$ value integrated over the Ge volume as the laser beam is moved along $y$. The integrated $c_y$ and $\Delta V_{ISHE}$ profiles are in excellent agreement, demonstrating that the ISHE signal delivers an accurate mapping of the spin density.

The measured voltage signal $\Delta V_{ISHE}$ strongly depends on the incident photon energy. When the photon energy $h\nu$ is at resonance with the Ge direct gap, only degenerate heavy holes and light holes states at the Γ point of the Brillouin zone participate to direct optical transitions to the conduction band. When the photon energy exceeds $E_d$ by more than 300 meV split-off states are also involved in the absorption process and the initial electron spin polarization is reduced [2],[17]. This trend is clearly shown in Fig. 4a, where $\Delta V_{ISHE}$, normalized to the incident power $W$, is displayed as a function of the photon energy. $\Delta V_{ISHE}/W$ is maximum for $h\nu$ = 0.8 eV and becomes negligible when $h\nu$ = 1.76 eV, in excellent agreement with theoretical calculations [17]. Figure 4b reveals instead that, up to $W$ = 400 µW, the dependence of the ISHE signal is linear with the number of initial spin-oriented electrons, ruling out thermally-induced effects.

We have also repeated the measurements described above by varying the stripe width from 4 µm down to 500 nm (see Fig. 4c). Interestingly, $\Delta V_{ISHE}$ follows the same behavior within the entire investigated width range, whereas the slight decrease of the signal dynamics when decreasing the stripe width can be rather attributed to the diffraction-limited spatial resolution of the setup. This is a direct consequence of the intrinsic nature of the process under discussion, which is not affected by the in-plane dimension of the nanostructures. This paves the possibility to engineer an ordered array of opposite spin densities within the nanometric scale, as demonstrated for the stripe set with 500 nm width.

In conclusion, we designed and experimentally tested a Ge-based opto-spintronic device in the infrared range that works as a spin photovoltaic cell, i.e. that provides the simultaneous generation of two opposite, spatially separated spin populations when circularly polarized light is absorbed. The optically induced spin voltage can be properly engineered by varying the stripe dimensions, while different combinations of semiconductor and patterned overlayer materials can be employed. The spatial control over the optical spin injection down to the nanometer scale allows the integrability of photonics with other spintronic devices, where the spin polarization of the carriers lies in the plane of the sample. The simplicity and versatility of the proposed device and the wide range of functionalities that can be exploited promise huge advantages in the design of a new class of spintronic devices.

**Acknowledgments**

The authors would like to thank Lamberto Duò for fruitful discussion. The CARIPLO foundation is acknowledged for partially funding this work through the NANOGAP (2010-0632) projects. The research leading to these results has received funding from the European Union's Seventh Framework Programme under grant agreement n° 613055.

**Figures**

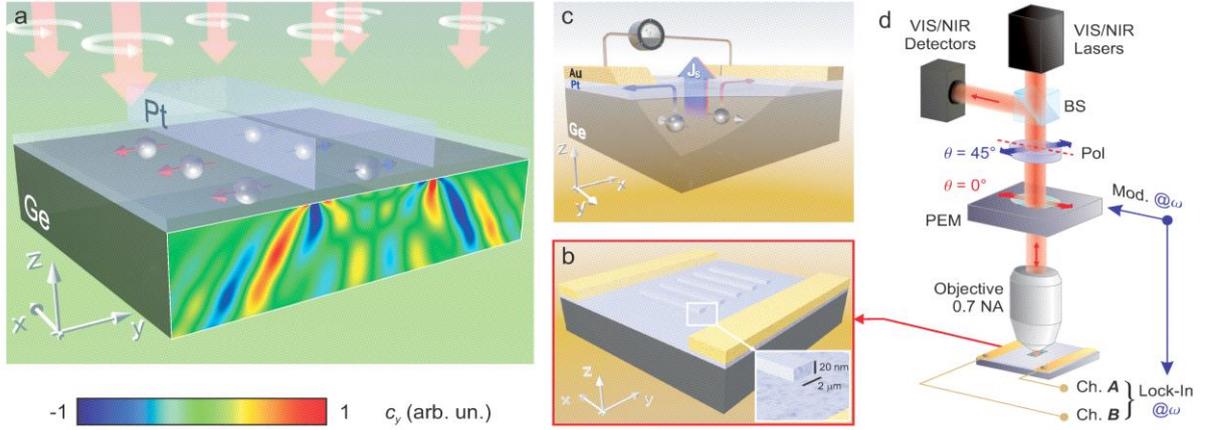

**Fig.1 a,** Illustration of the spin photovoltaic cell working principle: when a Pt microstripe is placed on the top of a Ge substrate, illumination at normal incidence ($z$-axis) with right-circularly polarized light ($hv = 0.8$ eV) allows simultaneously obtaining two symmetric spin densities with opposite spin inside Ge at correspondence with the stripe edges. The spin density projected along the $y$ axis is proportional to the Stokes parameter $c_y = 2\text{Im}(E_x E_z^*)$, being $E_x$ and $E_z$ the (complex) components of the electric field along the $x$ and $z$ axes,. **b,** A sketch of the sample geometry. **c,** ISHE geometry in the investigated sample: the photo-generated electrons with a spin polarization $P_y$ yield a spin current density $J_s$ diffusing along the $z$-axis from n-doped Ge to the Pt layer, where the ISHE provides the conversion of $J_s$ into the electromotive force $\Delta V_{ISHE}$, measured at the edges of the Pt layer at correspondence with the 50 nm-thick Au contacts. **d,** Confocal setup for spatially-resolved ISHE measurements: the collimated excitation beam enters the photo-elastic modulator (PEM), which modulates the light circular polarization at $\omega = 50$ kHz. The light beam is then focused by a microscope objective (numeric aperture, NA = 0.7) on the Pt microstructures. While a raster scan of the sample is performed, the differential voltage signal $\Delta V_{ISHE}$ is detected by a lock-in amplifier demodulating at $\omega$.

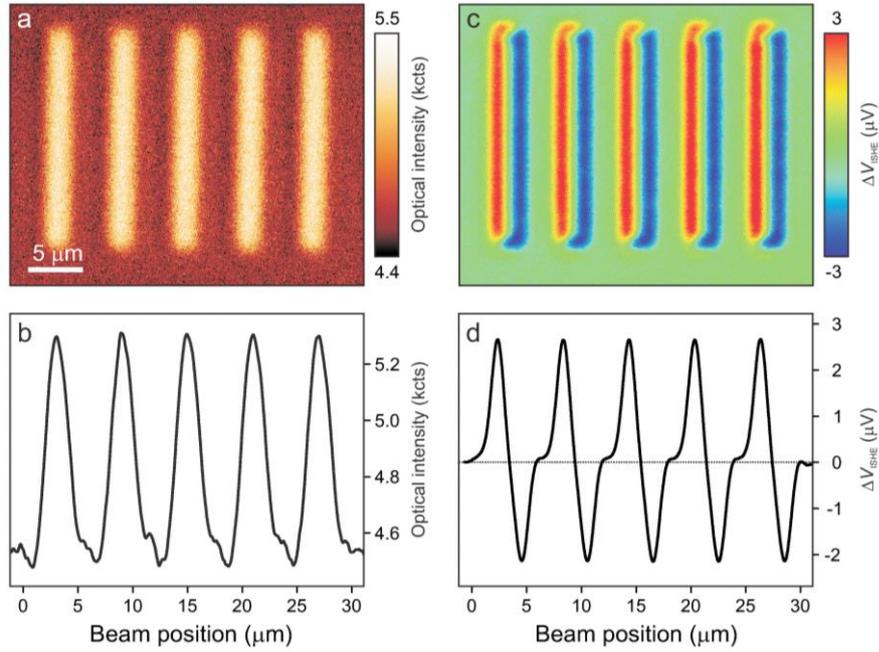

**Fig.2 a,** 2D optical reflectivity map of the nanostructured Pt/n-doped Ge sample, monitored by a InGaAs detector. The scanned area is 40×30 μm. **b,** Average of a set of 100 horizontal profiles taken across the map in **a** along the whole length of the stripes. **c,** 2D map of the differential voltage signal $\Delta V_{ISHE}$, measured with an excitation photon energy $h\nu = 0.8$ eV and incident power $W = 150$ μW. **d,** Average of a set of 100 horizontal profiles taken across the map in **c** along the whole length of the stripes in **c**.

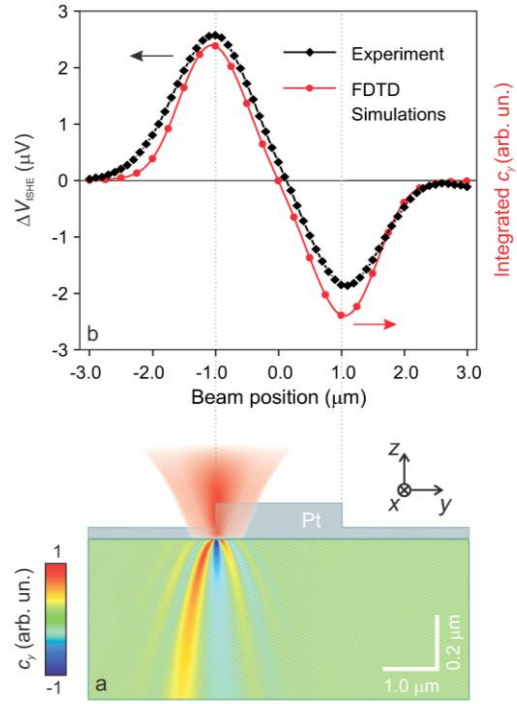

**Fig.3 a,** FDTD numerical simulation displaying the map of $c_y$ obtained when a Gaussian light beam ($h\nu$ = 0.8 eV, full width at half maximum $w \approx$ 1.5 μm) is shone on one edge of the Pt stripe. Note that $c_y$ display a net positive average value. The simulated region is 6×6×6 μm$^3$, with an adaptive mesh size down to 2 nm in proximity of the stripe edges. **b,** Comparison between the $\Delta V_{ISHE}$ average profile obtained across a 2 μm-wide Pt stripe and the corresponding calculated $c_y$ value integrated over the whole simulation region as a function of the beam position across the Pt stripe.

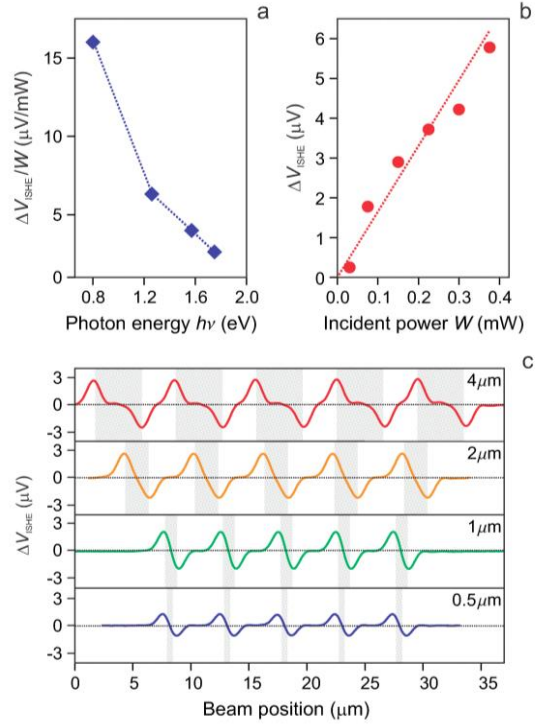

**Fig.4 a,** $\Delta V_{ISHE}$ normalized to the incident power $W$, as a function of the incident photon energy $h\nu$. $\Delta V_{ISHE}/W$ is maximum when $h\nu = 0.8$ eV, at resonance with the Ge direct bandgap at room temperature and follows the initial electron spin polarization curve discussed in the literature. **b,** $\Delta V_{ISHE}$ as a function of the incident power. The dotted line represents a linear fit. **c,** Dependence of the differential voltage signal $\Delta V_{ISHE}$ on the Pt stripe width for an incident photon energy $h\nu = 0.8$ eV and $W = 150$ µW. The gray bands indicate the stripe position.